\documentclass[showpacs]{revtex4}
\usepackage{amsmath}
\usepackage{amssymb}
\usepackage{graphicx}
\usepackage{color}
\usepackage{dcolumn}

\begin{document}

\newcolumntype{.}{D{.}{.}{-1}}

\title{
Mixed neutron-star-plus-wormhole systems: Linear stability analysis
}
\author{
Vladimir Dzhunushaliev,$^{1,2,3,4,}$
\footnote{
v.dzhunushaliev@gmail.com}
Vladimir Folomeev,$^{3,4,}$
\footnote{vfolomeev@mail.ru}
Burkhard Kleihaus,$^{4,}$
\footnote{b.kleihaus@uni-oldenburg.de } and
Jutta Kunz$^{4,}$
\footnote{jutta.kunz@uni-oldenburg.de}
}
\affiliation{
$^1$Department of Theoretical and Nuclear Physics, Kazakh National University, Almaty 050040, Kazhakhstan
\\
$^2$Institute for Basic Research,
Eurasian National University,
Astana 010008, Kazhakhstan
 \\ 
$^3$Institute of Physicotechnical Problems and Material Science of the NAS of the Kyrgyz Republic,  
265 a, Chui Street, Bishkek 720071,  Kyrgyzstan \\
$^4$Institut f\"ur Physik, Universit\"at Oldenburg, Postfach 2503,
D-26111 Oldenburg, Germany
}

\begin{abstract}
We consider configurations consisting of a neutron star
with a wormhole at the core. The wormhole is held open
by a ghost scalar field with a quartic coupling.
The neutron matter is described by a perfect fluid with
a polytropic equation of state.
We obtain static regular solutions for these systems.
A stability analysis, however, shows
that they are unstable with respect to linear perturbations.
\end{abstract}

\maketitle

\section{Introduction}
The geometrical model of electric charge suggested by Wheeler in the middle of the 1950s,
describes a tunnel connecting
two space-time regions that is filled by an electric field \cite{Wheeler:1955zz}.
This idea has stimulated many studies of solutions with
a nontrivial space-time topology  -- wormholes.
One of the most significant contributions in this area is the model
of a traversable Lorentzian wormhole,
suggested by Morris and Thorne  \cite{Thorne:1988}.
The traversability assumes that matter and radiation
can travel freely through the wormhole.
The key condition for the existence
of such a type of wormhole is the necessity to violate the weak/null energy conditions.
In Einstein gravity this means,
that the matter creating the wormhole must possess very exotic properties.

Thus the question arises whether such exotic forms of matter can exist in the Universe.
A strong argument in favor of such a possibility
is the observed accelerated expansion of the present Universe.
The cause of this acceleration is attributed to a repulsive
component of the Universe, the so-called dark energy.
Its amount is about 70\% of the total energy density of the Universe.
With its large negative pressure, dark energy causes the Universe
to expand increasingly fast, according to a power law or exponentially.
Moreover, astronomical observations  (see, e.g., Refs.~\cite{Tonry:2003zg,Alam:2003fg}
and more recent estimates \cite{Sullivan:2011kv})
indicate the possibility, that an even more exotic form of
energy exists in the Universe, called phantom dark energy.
The presence of such an energy assumes the violation of the weak/null energy conditions
and results in even faster acceleration.

If dark energy,  in one form or another,
does indeed exist in the Universe, this would give the basis for concluding that
localized compact objects consisting of dark energy might also exist.
Such objects could be so-called dark energy stars
\cite{Mazur:2004ku,Dymnikova:2004qg,Lobo:2005uf,DeBenedictis:2005vp,
DeBenedictis:2008qm,Gorini:2008zj,Gorini:2009em,Dzhunushaliev:2008bq,
Dzhunushaliev:2011ma,Folomeev:2011aa}
-- objects consisting of dark energy and possessing a trivial topology.
On the other hand, if phantom dark energy is present in the Universe
then it is possible
to imagine a situation where topologically nontrivial configurations~-- wormholes~-- can exist
\cite{kuhfittig,lxli,ArmendarizPicon:2002km,Sushkov:2002ef,Lemos:2003jb,lobo,sushkov}.
The possibility is not excluded that
the sizes of such configurations could be quite large
and comparable to those of various astrophysical objects -- such as
ordinary stars or neutron stars \cite{Kardashev:2006nj}.

One type of matter providing violation of the weak/null energy conditions
is the so-called ghost or phantom scalar field.
Although such fields generally suffer from quantum instabilities (their energy density  is unbounded from
below)~\cite{Carroll:2003st}, they have found quite a wide application in modeling the early and the present
accelerated Universe~\cite{Copeland:2006wr}.
With the opposite sign in front of its kinetic energy term,
such a field allows for solutions with nontrivial topology, including traversable wormholes.
The properties of such solutions as, for instance, their regularity and stability then depend
on the particular field employed.
Perhaps the simplest possibility to obtain wormhole-like solutions
is to consider massless scalar fields~\cite{Bronnikov:1973fh,Ellis:1973yv}.
However, as recently shown, such wormhole configurations
are unstable with respect to linear \cite{Gonzalez:2008wd,Bronnikov:2011if}
and nonlinear perturbations \cite{Gonzalez:2008xk}.

When stepping beyond the bounds of Einstein gravity and considering
the issue of the stability of wormhole solutions
within the general framework of scalar-tensor theories
with massless nonminimally coupled scalar fields
it is seen that such solutions
are also unstable under linear perturbations
\cite{bronn3,bronn4,Bronnikov:2002qx,Bronnikov:2005an}.

In order to obtain stable solutions,
the study of configurations supported by a ghost scalar field
with a self-interaction potential might seem promising.
For wormholes such potentials were considered
in Refs.~\cite{Kodama:1978dw,Kodama:1979}.
In particular,  here a ghost scalar field with a quartic coupling was employed,
and it was concluded that
regular, stable solutions with topologically nontrivial (wormhole-like) geometry exist.
A similar conclusion was reached in Ref.~\cite{Dzhunushaliev:2010bv}
for wormholes with a sine-Gordon ghost scalar field.

Here our objective is the study of configurations with nontrivial topology
consisting both of ordinary and exotic matter.
We have considered such mixed gravitating systems before in
Refs.~\cite{Dzhunushaliev:2011xx,Dzhunushaliev:2012ke},
where we suggested the possible existence of configurations
consisting of a traversable wormhole
(supported by a massless ghost scalar field)
filled by a perfect polytropic fluid.
We have shown that static, regular solutions can indeed be constructed,
which describe such mixed star-plus-wormhole systems.
These possess new physical properties
which distinguish them from ordinary stars.

In Ref.~\cite{Dzhunushaliev:2011xx} we made some preliminary estimates concerning
the stability of such mixed configurations with respect to linear perturbations.
However,  our analysis was incomplete since it was performed
only in the external region of the star.
On the other hand, bearing in mind that a wormhole supported by a massless scalar field
without ordinary matter is unstable \cite{Gonzalez:2008wd,Bronnikov:2011if},
we may naively expect that adding ordinary matter to such a configuration
will not lead to the stabilization of the system.
One reason could be that the main contribution to the energy density
near the throat is coming from the scalar field,
as our studies of mixed configurations performed
in Refs.~\cite{Dzhunushaliev:2011xx,Dzhunushaliev:2012ke} indicated.

As mentioned above, in Refs.~\cite{Gonzalez:2008wd,Bronnikov:2011if}
the question of the stability of wormhole-like solutions
with a massless ghost scalar field was clarified.
In particular, it was shown that when allowing for perturbations of the throat radius
the solutions are unstable with respect to spherically symmetric perturbations.
Recent investigations in Ref.~\cite{Bronnikov:2012ch} showed
that wormhole-like solutions remain unstable
for special choices of the scalar field potential.

Bearing all this in mind, in the present paper we reanalyze
the stability of wormhole solutions for a ghost scalar field
with a quartic coupling,
first addressed in Ref.~\cite{Kodama:1979}.
Unlike those pioneering calculations, we now allow for perturbations of the throat radius.
Subsequently, we perform such a linear stability analysis
for the mixed configurations, consisting
of a wormhole and both ordinary matter and a ghost scalar field
with a quartic coupling.

The paper is organized as follows. In Sec.~\ref{gen_equations}
the general set of equations is derived for configurations
consisting of a neutron star with a wormhole at the core, where
the neutron matter is modeled by a perfect fluid with a polytropic equation of state.
In Sec.~\ref{static_solutions_mex}, we present numerically obtained
static solutions for such topologically nontrivial configurations supported by
a ghost scalar field with a quartic coupling.
In Sec.~\ref{stab_anal}, a linear stability analysis is performed for these solutions.
Finally, in Sec.~\ref{conclus} our results are summarized.

\section{Static configurations}

\subsection{General equations}

\label{gen_equations}
We consider a model of a gravitating ghost scalar field
in  the  presence  of  a perfect fluid.
The Lagrangian for this system is chosen as
\begin{equation}
\label{lagran_wh_star_poten}
L=-\frac{c^4}{16\pi G}R-\frac{1}{2}\partial_{\mu}\varphi\partial^{\mu}\varphi -V(\varphi)+L_m.
\end{equation}
Here $\varphi$ is the ghost scalar field with the potential
 $V(\varphi)$, and $L_m$ is the Lagrangian of the perfect isotropic fluid
(where isotropic means that the radial and the tangential pressure
of the fluid agree),
which has the form $L_m=p$ \cite{Stanuk1964,Stanuk}.
Using this Lagrangian, the corresponding energy-momentum tensor can be presented as
 \begin{equation}
\label{emt_wh_star_poten}
T_i^k=(\varepsilon+p)u_i u^k-\delta_i^k p-\partial_{i}\varphi\partial^{k}\varphi
-\delta_i^k\left[-\frac{1}{2}\partial_{\mu}\varphi\partial^{\mu}\varphi-V(\varphi)\right],
\end{equation}
where $\varepsilon$ and $p$ are the energy density
and the pressure of the fluid, and $u^i$ is the four-velocity.
The metric can be taken in the general form
 \begin{equation}
\label{metric_gen}
ds^2=e^{\nu}(dx^0)^2-e^{\lambda} dr^2-e^{\mu} d\Omega^2,
\end{equation}
where $\nu,\lambda$, and $\mu$ are functions of the radial coordinate $r$
and the time coordinate  $x^0=c\, t$,
and $d\Omega^2$ is the metric on the unit two-sphere.

In considering equilibrium wormhole-like configurations,
it is convenient  to use the polar Gaussian coordinates
 \begin{equation}
\label{metric_wh_poten}
ds^2=e^{\nu}(dx^0)^2-dr^2-e^{\mu} d\Omega^2,
\end{equation}
where now $\nu$ and $\mu$ are functions of $r$ only.
Introducing the new function $R$ defined by $e^{\mu}=R^2$, the
$(_0^0)$, $(_1^1)$, and $(_2^2)$ components of the Einstein equations with metric
 \eqref{metric_wh_poten}  take the form
\begin{eqnarray}
\label{Einstein-00_poten}
&&-\left[2\frac{R^{\prime\prime}}{R}+\left(\frac{R^\prime}{R}\right)^2\right]+\frac{1}{R^2}
=\frac{8\pi G}{c^4} T_0^0=
\frac{8\pi G}{c^4}\left[ \varepsilon-\frac{1}{2}\varphi^{\prime 2}+V(\varphi)\right],
 \\
\label{Einstein-11_poten}
&&-\frac{R^\prime}{R}\left(\frac{R^\prime}{R}+\nu^\prime\right)+\frac{1}{R^2}
=\frac{8\pi G}{c^4} T_1^1=
\frac{8\pi G}{c^4}\left[- p+\frac{1}{2}\varphi^{\prime 2}+V(\varphi)\right],
\\
\label{Einstein-22_poten}
&&\frac{R^{\prime\prime}}{R}+\frac{1}{2}\frac{R^\prime}{R}\nu^\prime+
\frac{1}{2}\nu^{\prime\prime}+\frac{1}{4}\nu^{\prime 2}
=-\frac{8\pi G}{c^4} T_2^2=
\frac{8\pi G}{c^4}\left[ p+\frac{1}{2}\varphi^{\prime 2}-V(\varphi)\right],
\end{eqnarray}
where the prime denotes  differentiation with respect to $r$.

The equation for the scalar field $\varphi$ resulting from the Lagrangian \eqref{lagran_wh_star_poten} is
\begin{equation}
\label{sf_eq_gen}
\frac{1}{\sqrt{-g}}\frac{\partial}{\partial x^i}\left(\sqrt{-g}g^{ik}\frac{\partial \varphi}{\partial x^k}\right)=
\frac{d V}{d \varphi}.
\end{equation}
Using the metric \eqref{metric_wh_poten}, this equation gives
\begin{equation}
\label{sf_poten}
\varphi^{\prime\prime}+\left(\frac{1}{2}\nu^\prime+2\frac{R^\prime}{R}\right)\varphi^\prime=
-\frac{d V}{d \varphi}.
\end{equation}
Not all of the Einstein field equations are independent because of the
conservation of energy and momentum,
$T^k_{i;k}=0$. Taking the $i=1$ component of this equation gives
\begin{equation}
\label{conserv_1}
\frac{d T^1_1}{d r}+
\frac{1}{2}\left(T_1^1-T_0^0\right)\nu^\prime+2\frac{R^\prime}{R}\left[T_1^1-\frac{1}{2}\left(T^2_2+T^3_3\right)\right]=0.
\end{equation}
Taking into account the expressions
$$
T_2^2=T_3^3=- p-\frac{1}{2}\varphi^{\prime 2}+V(\varphi),
$$
and also Eq.~\eqref{sf_poten}, we obtain from Eq.~\eqref{conserv_1}
\begin{equation}
\label{conserv_2}
\frac{d p}{d r}=-\frac{1}{2}(\varepsilon+p)\frac{d\nu}{d r}.
\end{equation}

To model the matter filling the wormhole, it is
necessary to choose an appropriate equation of state. In doing so,
we proceed from the assumption that our configuration is essentially a
relativistic object, where the wormhole is filled with relativistic matter
having a pressure comparable with its energy density.
For this kind of matter we choose neutron matter.
In much of the literature neutron matter is described
by more or less conventional equations of state,
reflecting general properties of neutron matter at high densities and pressures.
Various forms of such equations of state can be found, for instance,
in Refs.~\cite{Oppen1939,Cameron1959,DAl1985,Haensel:2004nu}.

Since in the present paper we consider only general properties of mixed
neutron-star-plus-wormhole systems, we restrict ourselves to
a simplified variant of the equation of state,
where a more or less realistic neutron matter equation of state
is approximated in the form of a polytropic equation of state.
Namely, we employ the following parametric relation between
the pressure and the energy density of the fluid:
$$
\varepsilon=n_b m_b c^2
+ \frac{p}{\gamma-1} , \quad
p=k c^2 n_{b}^{(ch)} m_b \left(\frac{n_b}{n_{b}^{(ch)}}\right)^\gamma,
$$
where $n_{b}$ is the baryon number density,
$n_{b}^{(ch)}$ is some characteristic value of $n_{b}$,
$m_b$ is the baryon mass,
and $k$ and $\gamma$ are parameters
whose values depend on the properties of the neutron matter.

It is convenient to rewrite the above equation of state (EOS)
in the form
\begin{equation}
\label{eqs_NS_WH}
p=K \rho_{b}^{1+1/n}, \quad \varepsilon = \rho_b c^2 +n p,
\end{equation}
with the constant $K=k c^2 (n_{b}^{(ch)} m_b)^{1-\gamma}$,
the polytropic index $n=1/(\gamma-1)$,
and $\rho_b=n_{b} m_b$ denotes the rest-mass density
of the neutron fluid.

Setting
$m_b=1.66 \times 10^{-24}\, \text{g}$
and $n_{b}^{(ch)} = 0.1\, \text{fm}^{-3}$,
we consider below configurations with
$k=0.1$ and $\gamma=2$   \cite{Salg1994},
corresponding to a gas of baryons interacting
via a vector-meson field, as described by Zel'dovich \cite{Zeld1961,Zeld}
(see also Ref.~\cite{Tooper2}
where relativistic configurations
with such an equation of state were considered).
In Ref.~\cite{Dzhunushaliev:2012ke} we have already considered a similar mixed system
consisting of a massless ghost scalar field and a neutron fluid with an EOS
in the form of Eq.~\eqref{eqs_NS_WH}. Here we extend those results to the case where
the scalar field potential is present.

Introducing the new variable $\theta$ \cite{Zeld},
\begin{equation}
\label{theta_def}
\rho_b=\rho_{b c} \theta^n~,
\end{equation}
where $\rho_{b c}$ is the density of the neutron fluid at the
wormhole throat
(or, in other words, at the core of the configuration),
we may rewrite the pressure and the energy density,
Eq.~\eqref{eqs_NS_WH}, in the form
\begin{equation}
\label{pressure_fluid_theta}
p=K\rho_{b c}^{1+1/n} \theta^{n+1}, \quad
\varepsilon =  \left( \rho_{b c} c^2 +
  n K \rho_{b c}^{1 + {1}/{n} } \theta \right) \theta^n.
\end{equation}
Making use of this expression,
we obtain
for the internal region with $\theta \ne 0$
from Eq.~\eqref{conserv_2}
\begin{equation}
\label{conserv_3}
2\sigma(n+1)\frac{d\theta}{d r}=
-\left[1+\sigma(n+1) \theta\right]\frac{d\nu}{dr},
\end{equation}
where $\sigma=K \rho_{b c}^{1/n}/c^2=p_c/(\rho_{b c} c^2)$ is a constant,
related to the pressure $p_c$ of the fluid at the wormhole throat.
This equation may be integrated to
give in the internal region with $\theta \ne 0$ the metric function $e^{\nu}$
in terms of $\theta$,
\begin{equation}
\label{nu_app}
e^{\nu}=e^{\nu_c}\left[\frac{1+\sigma (n+1)}{1+\sigma (n+1)\theta}\right]^{2},
\end{equation}
and $e^{\nu_c}$ is the value of $e^{\nu}$ at the throat where $\theta=1$.
The integration constant $\nu_c$ is fixed
by requiring that the space-time is asymptotically flat,
i.e., $e^{\nu}=1$ at infinity.

Thus we have three unknown functions~-- $R, \theta$, and $\varphi$~-- for which there are four equations,
\eqref{Einstein-00_poten}-\eqref{Einstein-22_poten} and
\eqref{sf_poten} (only three of which are independent), and also the relation
 \eqref{nu_app}.
For the numerical calculations it is convenient to rewrite these equations in terms of dimensionless variables.
Since in Sec.~\ref{static_solutions_mex} we will consider
a particular case, where the scalar field $\varphi$ is equal to zero
at the wormhole throat, but its derivative is nonzero, we can introduce dimensionless variables as follows.
The potential can be expanded  in the neighborhood of the throat as
$$
\varphi \approx \varphi_1 r +\frac{1}{6}\varphi_3 r^3,
$$
where $\varphi_1$ is the derivative at the throat, the square of which corresponds to the ``kinetic'' energy of
scalar field. Then, it is convenient to use new dimensionless variables
expressed in units of $\varphi_1^2$.
 Namely, introducing
\begin{equation}
\label{dimless_xi_v}
\xi=\frac{r}{L}, \quad \Sigma=\frac{R}{L},
\quad \phi(\xi)=\frac{\sqrt{8\pi G}}{c^2}\,\varphi(r),
\quad \text{where} \quad L=\frac{c^2}{\sqrt{8\pi G}\varphi_1},
\end{equation}
with $L$
having  dimensions of  length, one can rewrite
Eqs.~\eqref{Einstein-00_poten}-\eqref{Einstein-22_poten} and
\eqref{sf_poten} in the form
\begin{eqnarray}
\label{Einstein-00_dmls}
&&-\left[2\frac{\Sigma^{\prime\prime}}{\Sigma}+\left(\frac{\Sigma^\prime}{\Sigma}\right)^2\right]+\frac{1}{\Sigma^2}
=B  (1+\sigma n \theta) \theta^n
-\frac{1}{2}\phi^{\prime 2}+\tilde{V},
 \\
\label{Einstein-11_dmls}
&&-\frac{\Sigma^\prime}{\Sigma}\left(\frac{\Sigma^\prime}{\Sigma}+\nu^\prime\right)+\frac{1}{\Sigma^2}
=-B \sigma \theta^{n+1}
+\frac{1}{2}\phi^{\prime 2}+\tilde{V},
\\
\label{Einstein-22_dmls}
&&\frac{\Sigma^{\prime\prime}}{\Sigma}+\frac{1}{2}\frac{\Sigma^\prime}{\Sigma}\nu^\prime+
\frac{1}{2}\nu^{\prime\prime}+\frac{1}{4}\nu^{\prime 2}
=
B \sigma \theta^{n+1}
+\frac{1}{2}\phi^{\prime 2}-\tilde{V},
\\
\label{sf_dmls}
&&\phi^{\prime\prime}+\left(\frac{1}{2}\nu^\prime+2\frac{\Sigma^\prime}{\Sigma}\right)\phi^\prime=
-\frac{d \tilde{V}}{d \phi},
\end{eqnarray}
where $\tilde{V}=V/\varphi_1^2$ is the dimensionless potential of the field, and
$B=(\rho_{b c} c^2)/\varphi_1^2$ is the dimensionless ratio of the fluid energy density to that
of the scalar field at the throat.

Thus, the static configurations under consideration are described by any
three equations from the system \eqref{Einstein-00_dmls}-\eqref{sf_dmls} together with Eq.~\eqref{conserv_3} or Eq.~\eqref{nu_app}.
Note, that in the case of $B=0$ we are dealing with a system consisting of
a pure scalar field configuration with no ordinary matter.

\subsection{Quartic potential
}
\label{static_solutions_mex}

In this section we discuss the numerical solutions of the set of equations  \eqref{conserv_3}, \eqref{Einstein-00_dmls}-\eqref{sf_dmls}.
We seek regular solutions of these equations
describing configurations with a finite mass.
In the case of a massless scalar field, it was shown in Refs.~\cite{Dzhunushaliev:2011xx,Dzhunushaliev:2012ke}
that static solutions for mixed systems consisting of a scalar field and a polytropic fluid
can indeed be obtained.

In the present paper
we consider the case where the potential has the well-known form of
the $\varphi^4$ theory. It was shown in
Refs.~\cite{Kodama:1978dw,Kodama:1979}  that a ghost field with such a potential
admits regular topologically nontrivial solutions.
Our aim here is to study the influence which the presence of a
polytropic fluid has on such solutions.

For our purpose, we choose the potential term in the form
 \cite{Kodama:1978dw,Kodama:1979}
$$
V=-\frac{1}{2}\left(\frac{m_\varphi}{f}\right)^2\left(1-f^2\varphi^2\right)^2,
$$
where $m_\varphi$ and $f$ are constants. Using the dimensionless variables \eqref{dimless_xi_v}, one can rewrite this potential as follows:
\begin{equation}
\label{poten_mex}
\tilde{V}=-\tilde{V}(0)\left(1-\Lambda^2\phi^2\right)^2,
\end{equation}
where $\Lambda^2=f^2 c^4/(8\pi G)$ is a dimensionless constant, and
 $\tilde{V}(0)=\left(m_\varphi/f\right)^2/(2\varphi_1^2)$ is the value of the potential at the local minimum, where
$\phi_{\text{min}}=0$. Maxima of this potential are located at the points
 $\phi_{\text{max}}=\pm 1/\Lambda$. Asymptotically, as
 $\xi \to \infty$, a regular solution must approach one of these maxima.
Note here that, in constrast to the case of usual (nonghost) scalar fields,
the existence of regular solutions is only possible, when the sign of the potential
is reversed with respect to the usual case.
This is the reason that we choose a potential of the form \eqref{poten_mex},
that is unbounded from below.

Equations \eqref{Einstein-00_dmls}-\eqref{sf_dmls} are to be solved
for given $\sigma$, $n$, and $B$,
subject to the boundary conditions at the core of the configuration $\xi=0$,
\begin{equation}
\label{bound_stat}
\theta(0) = 1, \quad \Sigma(0)=\Sigma_c, \quad \Sigma^\prime (0)=0,
\quad \nu(0)=\nu_c, \quad
\phi(0) =0, \quad \phi^\prime (0)=1.
\end{equation}
The quantity $\Sigma_c$ is the eigenparameter  of the system.
It is determined from the condition of obtaining
asymptotically vacuum solutions, where $\phi \to \pm 1/\Lambda$.
In turn, the value of the parameter  $B$ may be obtained
by expressing $\varphi_1$ in terms of $L$ from Eq.~\eqref{dimless_xi_v},
and this gives $B=8\pi G \rho_{b c} (L/c)^2$.
Thus, the value of $B$ is determined by the core density of the fluid and the choice
of the characteristic size $L$ of the configuration under consideration.
The case $B\to 0$ corresponds to the exclusion of the fluid from the system,
leaving only wormholes supported by a scalar field \cite{Kodama:1978dw,Kodama:1979}.

Let us emphasize again that, for a given value of $\Lambda$,
and thus a given theory,
there is only a single value of the wormhole size $\Sigma_c$
for which a regular solution exists [with the potential \eqref{poten_mex}]
 \cite{Kodama:1978dw,Kodama:1979}.
This situation differs from other variants of
bosonic configurations considered in the literature.
For example, in the case of systems
with self-interacting complex scalar fields considered in Refs.~\cite{Colpi:1986ye,Gleiser:1988rq},
the mass of the configurations, for a fixed coupling constant,
is a function of the central value of the scalar field,
and correspondingly of its central energy density.
In this case one can obtain the dependence of the mass on the central density which,
as in the case of configurations consisting only of ordinary (for instance, neutron) matter,
may be employed in considering the stability of the configurations
within the energy approach.
However, for the wormholes of Refs.~\cite{Kodama:1978dw,Kodama:1979}
such an analysis cannot be performed.

For the mixed configurations considered in this paper, where~--
besides the ghost scalar field with a $\phi^4$ potential~--
also ordinary matter is present,
the total mass of the configurations depends on the amount of fluid in the system,
and correspondingly on the fluid core density.
Then the possibility of considering the stability
of the configurations within the energy approach
appears again possible and will be discussed below.

Substituting the potential
 \eqref{poten_mex} into Eqs.~\eqref{conserv_3} and \eqref{Einstein-00_dmls}-\eqref{sf_dmls}
and using the boundary conditions \eqref{bound_stat},
we  seek a numerical solution of this set of equations.
In doing so, the configurations under consideration can be subdivided into two regions:
 (i) the internal one, where both the scalar field and the fluid are present;
 (ii) the external one, where only the scalar field is present.
Correspondingly, the solutions in the external region are obtained by using Eqs.~\eqref{Einstein-00_dmls}-\eqref{sf_dmls},
in which $\theta$ is set to zero.
The internal solutions must be matched with the external ones at the boundary of the fluid,
$\xi=\xi_b$, by equating the corresponding values of the functions $\phi, \Sigma, \nu$
and their derivatives.
The boundary of the fluid $\xi_b$ is defined by $p(\xi_b)=0$.
Knowledge of the asymptotic solutions in turn allows one
to determine the value of the integration constant $\nu_c$ at the throat,
proceeding from the requirement
of asymptotic flatness of the external solutions.

Let us now address the total mass of the configurations.
For the spherically symmetric metric \eqref{metric_wh_poten},
the mass $m(r)$ inside the radius $r$ can be defined as follows:
\begin{equation}
\label{mass_dm}
m(r)=\frac{c^2}{2 G}R_c+\frac{4\pi}{c^2}\int_{R_c}^{r} T_0^0 R^2   dR,
\end{equation}
where $R_c$ is the radius of the wormhole throat defined by $R_c=\text{min}\{R(r)\}$.
Without loss of generality, we can take this throat to occur at $r=0$.
Note, that for the total mass the upper limit of the integral is infinity,
since formally the energy density of the scalar field becomes equal to zero only asymptotically,
as $R\to \infty$. In practice, however, the scalar field decays exponentially fast.
Consequently, for the values of the parameters employed
all mass is concentrated within a size of order $L$.
Note also, that in evaluating the above integral it is necessary
to perform the calculations separately in the internal and external regions.

\begin{table}
 \caption{Characteristics of a set of configurations for EOS \eqref{eqs_NS_WH}.
The radius of the throat $R_{\text{th}}$,
the proper radius of the fluid $R_{\text{prop}}$,
and the gravitational radius $r_g$ (last column)
are given in kilometers.
The total mass $M$, the mass at the throat $M_{\text{th}}$,
the mass of the fluid $M_{\text{fl}}$,
and the external part of the mass of the scalar field $M_{\text{sfext}}$
are given in solar mass units.}
\vspace{.3cm}
\begin{tabular}{p{1.5cm}p{1.5cm}p{1.5cm}p{1.5cm}p{1.5cm}p{1.5cm}p{1.5cm}p{1.5cm}p{1.5cm}}
\hline \\[-5pt]
$\rho_{b c}, \text{g cm}^{-3}$ &  $ \hphantom{xx}R_{\text{th}},$ km &  $ \hphantom{}R_{\text{prop}},$ km & $\hphantom{xx}B$ &
 $\hphantom{xx} M/M_\odot$ &  $\hphantom{x} M_{\text{th}}/M_\odot$&  $\hphantom{x} M_{\text{fl}}/M_\odot$
&    $\hphantom{x} M_{\text{sfext}}/M_\odot$& $ \hphantom{xx} r_g,$ km \\[2pt]
\hline \\[-7pt]
\multicolumn{9}{c}{Without a wormhole} \\[2pt]
\end{tabular}
\begin{tabular}{p{1.5cm}........}
\hline \\[-15pt]
1.0$\times 10^{13}$&	-&	35.3940&	-&	0.2777&	-&	0.2777&		-&	0.8192\\
3.0$\times 10^{13}$&	-&	34.7518&	-&	0.7573&	-&	0.7573&		-&	2.2338\\
5.0$\times 10^{13}$&	-&	34.1519&	-&	1.1532&	-&	1.1532&		-&	3.4014\\
7.0$\times 10^{13}$&	-&	33.5906&	-&	1.4818&	-&	1.4818&		-&	4.3708\\
1.0$\times 10^{14}$&	-&	32.8113&	-&	1.8760&	-&	1.8760&		-&	5.5335\\
2.0$\times 10^{14}$&	-&	30.6535&	-&	2.6482&	-&	2.6482&		-&	7.8112\\
3.0$\times 10^{14}$&	-&	28.9877&	-&	2.9832&	-&	2.9832&		-&	8.7993\\
4.0$\times 10^{14}$&	-&	27.6575&	-&	3.1213&	-&	3.1213&		-&	9.2068\\
6.0$\times 10^{14}$&	-&	25.6722&	-&	3.1564&	-&	3.1564&		-&	9.3101\\
8.0$\times 10^{14}$&	-&	24.2625&	-&	3.0779&	-&	3.0779&		-&	9.0787\\
1.0$\times 10^{15}$&	-&	23.2159&	-&	2.9715&	-&	2.9715&		-&	8.7647\\
1.5$\times 10^{15}$&	-&	21.5291&	-&	2.7156&	-&	2.7156&		-&	8.0099\\
2.0$\times 10^{15}$&	-&	20.5768&	-&	2.5160&	-&	2.5160&		-&	7.4212\\
3.0$\times 10^{15}$&	-&	19.7367&	-&	2.2541&	-&	2.2541&		-&	6.6487\\

\hline
\multicolumn{9}{c}{$\phantom{\Big(}\Lambda=1, L=10 \,\text{km}$}\\[2pt]
\hline \\[-15pt]
1.0$\times 10^{13}$&	2.0075&	0.3165&	0.0002&	0.3823&	0.6806&0.0000&	-0.2520&	1.1277\\
5.0$\times 10^{13}$&	2.0078&	0.7822&	0.0009&	0.3827&	0.6807&	0.0002&	-0.1186&	1.1288\\
1.0$\times 10^{14}$&	2.0084&	1.2784&	0.0019&	0.3837&	0.6809&	0.0008&	-0.0366&	1.1318\\
3.0$\times 10^{14}$&	2.0134&	5.5204&	0.0056&	0.4078&	0.6826&	0.0231&	-0.0000&	1.2029\\
4.0$\times 10^{14}$&	2.0172&	13.1029&	0.0075&	0.5194&	0.6839&	0.1335&	-0.0000&	1.5321\\
5.0$\times 10^{14}$&	2.0219&	21.8576&	0.0093&	0.9602&	0.6855&	0.5731&	-0.0000&	2.8322\\
6.0$\times 10^{14}$&	2.0276&	24.5778&	0.0112&	1.4644&	0.6874&	1.0760&	-0.0000&	4.3195\\
7.0$\times 10^{14}$&	2.0343&	25.0595&	0.0130&	1.8199&	0.6897&	1.4300&	-0.0000&	5.3679\\
1.0$\times 10^{15}$&	2.0603&	24.0895&	0.0186&	2.2767&	0.6985&	1.8818&	-0.0000&	6.7153\\
1.5$\times 10^{15}$&	2.1272&	22.0069&	0.0279&	2.3277&	0.7212&	1.9217&	-0.0000&	6.8658\\
2.0$\times 10^{15}$&	2.2318&	20.2965&	0.0373&	2.1844&	0.7566&	1.7645&	-0.0000&	6.4432\\
3.0$\times 10^{15}$&	2.6474&	16.4199&	0.0559&	1.8366&	0.8976&	1.3854&	-0.0000&	5.4172\\
3.5$\times 10^{15}$&	3.0947&	13.4552&	0.0652&	1.7418&	1.0492&	1.2971&	-0.0000&	5.1377\\

\hline
\end{tabular}
\label{tab1}
\end{table}

\begin{figure}[h!]
\centering
  \includegraphics[height=10.8cm]{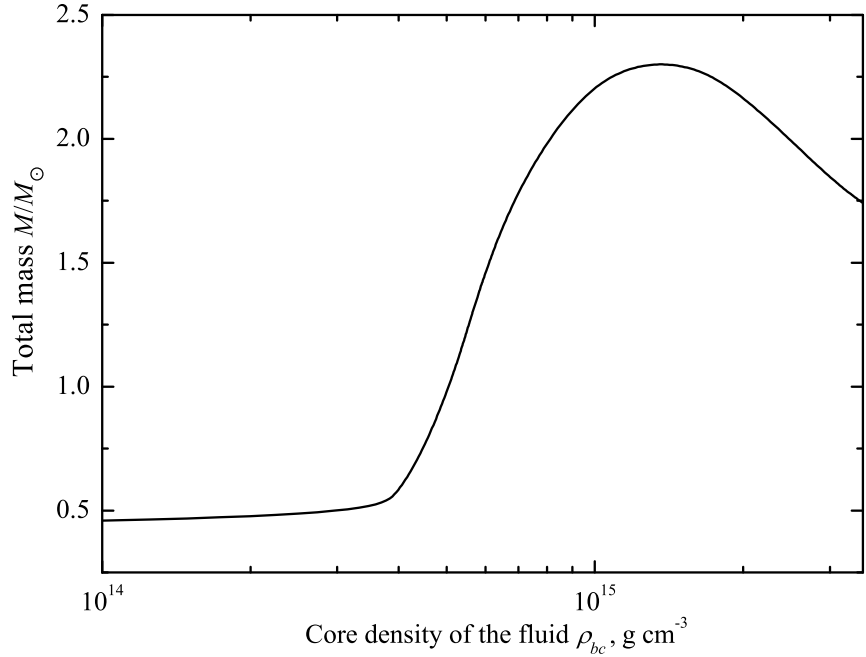}
\vspace{-1.cm}
\caption{Total mass of the configurations (in solar mass units), $M/M_\odot$,
versus the core density $\rho_{b c}$ (in units of  $\text{g cm}^{-3}$)
for EOS \eqref{eqs_NS_WH} with $k=0.1$ and $\gamma=2$ ($n=1$).
The characteristic size $L$ is taken as 10 km.
The value of $\Lambda$ is taken as 1.
Stable configurations should reside to the left of the first mass peak, if they would exist.
}
\label{mass_rho_c}
\end{figure}

\begin{figure}[h!]
\centering
  \includegraphics[height=10.8cm]{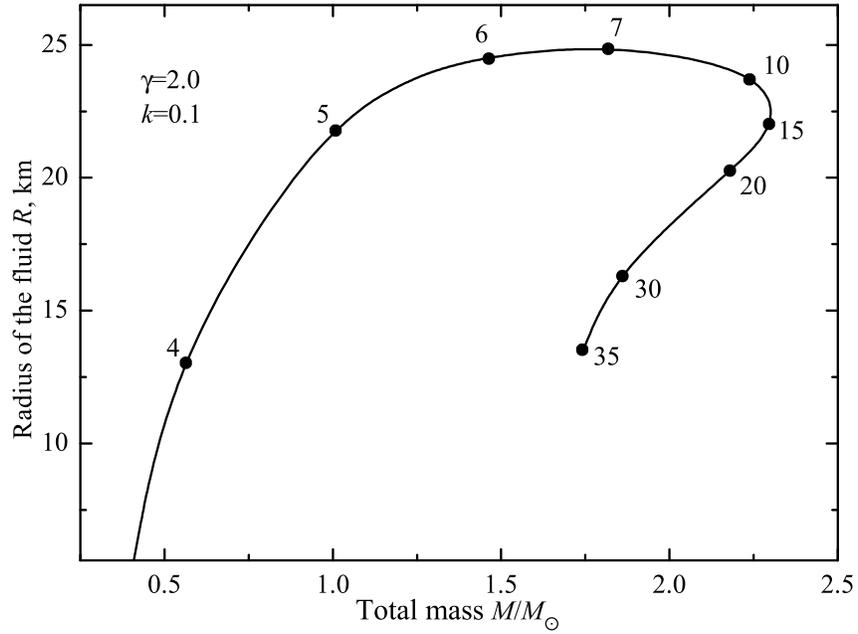}
\vspace{-1.2cm}
\caption{The mass-radius relation.
The points are labeled with the appropriate values of
the core density $\rho_{b c}$ (in units of $10^{14} \text{g cm}^{-3}$).
The tentative stability region is situated approximately in the range
$7\times 10^{14} \lesssim \rho_{b c} \lesssim 14\times 10^{14} \text{g cm}^{-3}$.
}
\label{M_R_relation}
\end{figure}

In the dimensionless variables \eqref{dimless_xi_v},  the expression
\eqref{mass_dm} takes the form
\begin{equation}
\label{mass_dmls}
m(\xi)=\frac{c^3}{2}\sqrt{\frac{B}{8\pi G^3 \rho_{b c}}}
\left\{\Sigma_c+
\int_{0}^{\xi} \Big[
B  (1+\sigma n \theta)\theta^n-\frac{1}{2}\phi^{\prime 2}+\tilde{V}
\Big]\Sigma^2 \frac{d\Sigma}{d\xi'}d\xi'
\right\},
\end{equation}
where the coefficient in front of the  curly brackets has the dimension of mass.

Using the data from Table \ref{tab1},
the numerical results are illustrated in Figs.~\ref{mass_rho_c} and \ref{M_R_relation}.
In performing the calculations,
we start from an initial configuration without ordinary matter (a pure wormhole).
Then, by gradually adding neutron matter to this system
(as expressed in the growth of $\rho_{b c}$ or, equivalently, of $B$)
we can monitor the behavior of the masses and sizes of the configurations.
The dependence of the total mass (in solar mass units)
on the core density of the fluid $\rho_{b c}$ (in grams per cubic centimeter)
is shown in Fig. \ref{mass_rho_c}.
As the core density decreases, i.e., as $B\to 0$,
the total mass of the configurations tends to the mass
of a pure wormhole without any fluid, $M_{\text{WH}}\approx 0.38 M_\odot$.

On the other hand, with increasing $\rho_{b c}$,
the total mass $M$ of the configurations rises
monotonically to a maximum and then decreases again.
This is typical for this type of configuration.
In Tooper's paper \cite{Tooper2}, for instance,
ordinary neutron stars were investigated in detail
for various values of the polytropic index $n$.
In particular, it was shown that the first peak in the mass
corresponds to the point dividing stable and unstable
neutron-star configurations.
This first mass peak is reached
at a critical value of the core density,
$\rho^{(\text{cr})}_{b c}$.

However, looking at Fig.~\ref{M_R_relation},
where the mass-radius relation for the configurations is presented,
one can see that as the core density of the fluid increases,
first a simultaneous growth of the size and the mass of the configurations takes place,
which is typical for unstable compact astrophysical systems.
In the interval
$7\times 10^{14} \text{g cm}^{-3} \lesssim \rho_{b c}
\lesssim  14\times 10^{14} \text{g cm}^{-3}$
the mass then continues to increase while the radius of the configurations decreases.
Such a behavior is typical for stable compact astrophysical systems.
Finally, for $\rho_{b c} \gtrsim 14\times 10^{14} \text{g cm}^{-3}$
the mass and the radius decrease simultaneously,
as  is characteristic for unstable configurations.

Thus, the presence of the wormhole has a remarkable effect on the configurations
for small central densities, since
for ordinary neutron stars (without a wormhole) the mass increases as the
radius decreases in this low-density range.
Consequently, ordinary neutron stars (without a wormhole) are stable in this
range, and remain stable all the way to the maximum of the mass.
Only beyond the associated critical value $\rho^{(\text{cr})}_{b c}$ would a star
become unstable (see, e.g., Ref.~\cite{Tooper2}).
The naive analogy with ordinary compact stars would then suggest,
that for star-plus-wormhole systems
instability occurs not only for $\rho_{b c}>\rho^{(\text{cr})}_{b c}$,
but also for small values of the core density $\rho_{b c}$.
Clearly, at this point a reliable stability analysis is called for.

\section{Linear stability analysis}
\label{stab_anal}

The simplest configurations with a nontrivial wormhole-like topology are
obtained by using a massless scalar field~\cite{Bronnikov:1973fh,Ellis:1973yv}.
While the first stability studies (see, e.g., Ref.~\cite{ArmendarizPicon:2002km})
did not find unstable modes,
recent work showed that these configurations
are linearly \cite{Gonzalez:2008wd,Bronnikov:2011if}
and nonlinearly  \cite{Shinkai:2002gv,Gonzalez:2008xk} unstable.
The reason for this discrepancy
is related to the fact that unlike
in Ref.~\cite{ArmendarizPicon:2002km}, in Refs.~\cite{Gonzalez:2008wd,Bronnikov:2011if,Gonzalez:2008xk}
the perturbations are not required to vanish at the throat.
In this case the resulting Schr\"{o}dinger-like equation used in the stability analysis
contains singularities in the corresponding effective potential that
does not allow one to perform a complete perturbation analysis over all space-time.
One way of solving this problem, used in Refs.~\cite{Gonzalez:2008wd,Bronnikov:2011if,Gonzalez:2008xk},
consists of regularizing the effective potential and solving the
regularized Schr\"{o}dinger-like equation.

In this section we perform a linear stability analysis  of the above static
solutions with the quartic potential \eqref{poten_mex}.
Also in this case it was shown, that the solutions
are stable against a special type of linear perturbation,
where oscillations of the throat radius are excluded \cite{Kodama:1979}.
Here we reanalyze the stability of these solutions,
by allowing for perturbations that are not required to vanish at the throat.

Subsequently, we consider the stability of star-plus-wormhole systems,
where besides the ghost scalar field also a polytropic fluid is present.
Our investigation is facilitated by the fact,
that both background solutions and perturbations decay exponentially fast,
when a $\phi^4$ potential considered~\cite{Kodama:1979}.

\subsection{General equations}

We now consider spherically symmetric perturbations of the above equilibrium configurations.
In obtaining the set of equations for the perturbations, we will neglect all quantities
which are of second and higher order.

For the energy-momentum tensor of the fluid we need
the components of the four-velocity in the metric \eqref{metric_gen}
\cite{Chandrasekhar:1964zz},
$$
u^0=e^{-\nu_0/2}, \quad u_0=e^{\nu_0/2}, \quad u^1=e^{-\nu_0/2} v, \quad u_1=-e^{\lambda_0-\nu_0/2} v,
$$
with the three-velocity
$$
v=\frac{d r}{d x^0} \ll 1 ~.
$$
The index 0 on the metric functions indicates the static, zeroth-order solutions of the Einstein equations.
The components of the energy-momentum tensor \eqref{emt_wh_star_poten}
then take the form
\begin{eqnarray}
\label{emt-00}
&&
T_0^0=\varepsilon-\frac{1}{2} e^{-\nu}\dot{\varphi}^2-\frac{1}{2} e^{-\lambda}\varphi^{\prime 2}+V(\varphi)
,\\
\label{emt-11}
&&
T_1^1=- p+\frac{1}{2} e^{-\nu}\dot{\varphi}^2+\frac{1}{2}e^{-\lambda} \varphi^{\prime 2}+V(\varphi),\\
\label{emt-10}
&&
T_0^1= (\varepsilon+p)u_0 u^1-\partial_0\varphi \partial^1\varphi=
(\varepsilon_0+p_0)v+ e^{-\lambda}\dot{\varphi}\,\varphi^\prime,\\
\label{emt-22}
&&
T_2^2=T_3^3=- p+\frac{1}{2} e^{-\nu}\dot{\varphi}^2-\frac{1}{2} e^{-\lambda}\varphi^{\prime 2}+V(\varphi).
\end{eqnarray}
In the above equations, the prime and dot denote differentiation with respect to $r$ and $x^0$,
respectively.

Now we consider perturbations of the static solutions.
Let us denote by $y$ any one of the functions $\nu, \lambda, \mu, \varepsilon, p$, or $\varphi$.
Then we assume that $y$ is of the form
\begin{equation}
\label{perturbations}
y=y_0+y_p ~,
\end{equation}
where the index 0 again refers to the static, zeroth-order solutions,
and the index $p$ indicates the perturbation.

Substituting these expressions into the $(_0^0)$, $(_1^1)$, and $(_2^2)$ components of the Einstein equations,
written in the metric~\eqref{metric_gen},
\begin{eqnarray}
\label{Einstein-00gen}
G_0^0&=&-e^{-\lambda}\left(\mu^{\prime\prime}+\frac{3}{4}\mu^{\prime 2}-\frac{1}{2}\mu^{\prime}\lambda^{\prime}\right)+
\frac{1}{4}e^{-\nu}\left(\dot{\mu}^2+2\dot{\mu}\dot{\lambda}\right)+e^{-\mu}=\frac{8\pi G}{c^4}T_0^0
,\\
\label{Einstein-11gen}
G_1^1&=&-\frac{1}{4}e^{-\lambda}\left(\mu^{\prime 2}+2\mu^{\prime}\nu^{\prime}\right)+
e^{-\nu}\left(\ddot{\mu}-\frac{1}{2}\dot{\mu}\dot{\nu}+\frac{3}{4}\dot{\mu}^2\right)+e^{-\mu}=\frac{8\pi G}{c^4}T_1^1~,\\
\label{Einstein-22gen}
G_2^2&=&\frac{1}{4}e^{-\lambda}\left(\lambda^\prime \mu^\prime-2\mu^{\prime\prime}-\mu^{\prime 2}-
\nu^\prime \mu^\prime+\lambda^\prime \nu^\prime-2\nu^{\prime\prime}-\nu^{\prime 2}\right) \nonumber\\
&+&
\frac{1}{4}e^{-\nu}\left(\dot{\lambda}\dot{\mu}+\dot{\mu}^2-\dot{\nu}\dot{\mu}+2\ddot{\mu}-
\dot{\lambda}\dot{\nu}+2\ddot{\lambda}+\dot{\lambda}^2\right)=\frac{8\pi G}{c^4}T_2^2~,
\end{eqnarray}
we find to linear order in $y_p$
\begin{eqnarray}
\label{Einstein-00pert}
&&
e^{-\lambda_0}\left[
\mu_p^{\prime\prime}+\frac{3}{2}\mu_0^\prime \mu_p^\prime-
\frac{1}{2}\left(\mu_0^\prime\lambda_p^\prime+\lambda_0^\prime \mu_p^\prime\right)-
\lambda_p\left(\mu_0^{\prime\prime}+\frac{3}{4}\mu_0^{\prime 2}-\frac{1}{2}\mu_0^\prime\lambda_0^\prime\right)
\right]+e^{-\mu_0}\mu_p \nonumber \\
&&=-\frac{8\pi G}{c^4}
\left[
\varepsilon_p-e^{-\lambda_0}\varphi_0^\prime\left(\varphi_p^\prime-\frac{1}{2}\varphi_0^\prime \lambda_p\right)+V_p
\right]
,\\
\label{Einstein-11pert}
&&
\frac{1}{2}e^{-\lambda_0}\left[
 \left(\nu_p^\prime+\mu_p^\prime\right)\mu_0^\prime+\nu_0^\prime \mu_p^\prime-
 \lambda_p\left(\frac{1}{2}\mu_0^{\prime 2}+\mu_0^\prime \nu_0^\prime\right)
\right]-e^{-\nu_0}\ddot{\mu}_p+e^{-\mu_0}\mu_p \nonumber \\
&&=-\frac{8\pi G}{c^4}
\left[
-p_p+e^{-\lambda_0}\varphi_0^\prime\left(\varphi_p^\prime-\frac{1}{2}\varphi_0^\prime \lambda_p\right)+V_p
\right],\\
\label{Einstein-22pert}
&&
\mu_p^{\prime\prime}+\nu_p^{\prime\prime}+\mu_0^\prime\left(\mu_p^\prime+\frac{1}{2}\nu_p^\prime-\frac{1}{2}\lambda_p^\prime\right)+
\nu_0^\prime\left(\frac{1}{2}\mu_p^\prime-\frac{1}{2}\lambda_p^\prime+\nu_p^\prime\right)
-\lambda_p\left[\mu_0^{\prime\prime}+\nu_0^{\prime\prime}+
\frac{1}{2}\left(\mu_0^{\prime 2}+\nu_0^{\prime 2}+\mu_0^\prime \nu_0^\prime\right)\right]
\nonumber \\
&&-e^{-\nu_0}(\ddot{\mu_p}+\ddot{\nu_p})=
-\frac{16\pi G}{c^4}\left(-p_p-e^{-\lambda_0}\varphi_0^\prime\left(\varphi_p^\prime-\frac{1}{2}\varphi_0^\prime \lambda_p\right)+V_p\right),
\end{eqnarray}
where $V_p=[\partial_\varphi V]_0 \,\varphi_p$. Next, from the (1-0)  component of the Einstein
equations,
$$
G^1_0=\frac{1}{2}e^{-\lambda}\left[
2\dot{\mu}^\prime-\dot{\lambda}\mu^\prime+\dot{\mu}\left(\mu^\prime-\nu^\prime\right)
\right]=\frac{8\pi G}{c^4}T^1_0
$$
we have to linear order
\begin{equation}
\label{Einstein-10pert}
\frac{1}{2}e^{-\lambda_0}\left[
2\dot{\mu_p}^\prime-\dot{\lambda_p}\mu_0^\prime+\dot{\mu_p}\left(\mu_0^\prime-\nu_0^\prime\right)
\right]=\frac{8\pi G}{c^4}\left[
\left(\varepsilon_0+p_0\right)v+e^{-\lambda_0}\dot{\varphi_p}\,\varphi_0^\prime
\right].
\end{equation}
Now we introduce a ``Lagrangian displacement'' $\zeta$ with respect to  $x^0$ \cite{Chandrasekhar:1964zz},
$$
v=\frac{\partial \zeta}{\partial x^0}\,.
$$
Then Eq.~\eqref{Einstein-10pert} can be integrated to give
\begin{equation}
\label{lambda_pert}
\lambda_p=\mu_p+\frac{2}{\mu_0^\prime}\left\{
\mu_p^\prime-\frac{1}{2}\mu_p\nu_0^\prime-\frac{8\pi G}{c^4}e^{\lambda_0}
\left[\left(\varepsilon_0+p_0\right)\zeta+e^{-\lambda_0}\varphi_0^\prime \varphi_p\right]
\right\}.
\end{equation}

In turn, the $i=1$ component of the
law of conservation of energy and momentum, $T^k_{i;k}=0$, gives
\begin{equation}
\label{conserv_osc}
\frac{\partial T^0_1}{\partial x^0}+\frac{\partial T^1_1}{\partial r}+\frac{1}{2}\left(\dot{\nu}+\dot{\lambda}
+2\dot{\mu}\right)T^0_1+
\frac{1}{2}\left(T_1^1-T_0^0\right)\nu^\prime+\mu^\prime\left[T_1^1-\frac{1}{2}\left(T^2_2+T^3_3\right)\right]=0.
\end{equation}
Substituting here the components \eqref{emt-00}-\eqref{emt-22}, we find
to linear order
\begin{align}
\label{conserv_osc_pert_gen}
\begin{split}
&-e^{\lambda_0-\nu_0}\left[(\varepsilon_0+p_0)\dot{v}+e^{-\lambda_0}\varphi_0^{\prime}\ddot{\varphi}_p\right]-
\frac{\partial p_p}{\partial r}+\frac{\partial V_p}{\partial r}
 \\
&+e^{-\lambda_0}\left\{
\Big(\varphi_0^{\prime\prime}-\lambda_0^{\prime}\varphi_0^{\prime}\Big)
\Big(\varphi_p^\prime-\frac{1}{2}\varphi_0^\prime \lambda_p\Big)+
\varphi_0^\prime\left[\varphi_p^{\prime\prime}-\frac{1}{2}
\left(\varphi_0^{\prime\prime}\lambda_p+\varphi_0^{\prime}\lambda_p^\prime\right)\right]
\right\}\\
& -\frac{1}{2}(\varepsilon_p+p_p)\nu_0^\prime-
\frac{1}{2}(\varepsilon_0+p_0)\nu_p^\prime\\
&+e^{-\lambda_0}\left[\frac{1}{2}\varphi_0^{\prime 2}\nu_p^\prime+\varphi_0^\prime
\left(\varphi_p^\prime-\frac{1}{2}\varphi_0^\prime \lambda_p\right)\nu_0^\prime\right]+
e^{-\lambda_0}\varphi_0^\prime\left[\varphi_0^\prime \mu_p^\prime+2\mu_0^\prime
\left(\varphi_p^\prime-\frac{1}{2}\varphi_0^\prime \lambda_p\right)
\right]=0.
\end{split}
\end{align}

The perturbed scalar field equation is found from Eq.~\eqref{sf_eq_gen},
and is given to linear order by
\begin{align}
\label{phi_pert_gen}
\begin{split}
&\varphi_p^{\prime\prime}- e^{\lambda_0-\nu_0}\ddot{\varphi}_p+
\frac{1}{2}\left(\nu_0^\prime-\lambda_0^\prime+2\mu_0^\prime\right)\varphi_p^\prime+
\frac{1}{2}\left(\nu_p^\prime-\lambda_p^\prime+2\mu_p^\prime\right)\varphi_0^\prime\\
&=-e^{\lambda_0}\left\{
\varphi_p \left[ \frac{d^2 V}{d\varphi^2}\right]_{\varphi=\varphi_0}+
\lambda_p\left[\frac{d V}{d\varphi}\right]_{\varphi=\varphi_0}
\right\}.
\end{split}
\end{align}

Thus, we have a set of five general linear equations~-- \eqref{Einstein-00pert}, \eqref{Einstein-11pert},
\eqref{lambda_pert}, \eqref{conserv_osc_pert_gen}, and \eqref{phi_pert_gen}~-- for the perturbations
$\theta_p, \nu_p, \lambda_p, \mu_p$, and $\varphi_p$.

\subsection{Quartic potential}
\label{stab_anal_mex}

Now we employ the set of equations obtained in the previous subsection to study the stability of the
static solutions considered in Sec.~\ref{static_solutions_mex}.
In this case the potential $V$ is given by Eq.~\eqref{poten_mex},
and the equation of state of the fluid is defined by Eq.~\eqref{pressure_fluid_theta}.
From these expressions one finds that the perturbed components of the pressure $p_p$ and the
energy density $\varepsilon_p$ are
\begin{equation}
\label{pres_energ_pert}
p_p=K (n+1) \rho_{b c}^{1+1/n}\theta_0^n \theta_p, \quad
\varepsilon_p=n \rho_{b c} c^2 \left[\frac{1}{\theta_0}+\sigma(n+1)\right]\theta_0^n \theta_p.
\end{equation}
In turn, the static components are
\begin{equation}
\label{pres_energ_non_pert}
p_0=K  \rho_{b c}^{1+1/n}\theta_0^{n+1}, \quad
\varepsilon_0= \rho_{b c} c^2 \left(1+\sigma n \theta_0\right)\theta_0^n.
\end{equation}

To proceed with the stability analysis
we now assume that the harmonic perturbations have the following time dependence:
\begin{equation}
\label{harmonic}
y_p(x^0,\xi) = \bar{y}_p(\xi) e^{i\omega x^0}~,
\end{equation}
where the functions $\bar{y}_p(\xi)$ depend only on the spatial coordinate $\xi$.
For convenience, we hereafter drop the bar.

Let us now consider the gauge freedom of the problem.
First, we have the freedom of choosing the radial coordinate $r$.
Here our choice has been to set $\lambda_0=0$.
Second, we can make a gauge choice for the metric perturbations $\nu_p, \lambda_p, \mu_p$.
In particular, we may impose a relation among the metric perturbations.
We are guided by the perturbed scalar field equation \eqref{phi_pert_gen},
which contains the term $\left(\nu_p^\prime-\lambda_p^\prime+2\mu_p^\prime\right)\varphi_0^\prime$.
This equation is considerably simplified by the gauge choice
$$\nu_p-\lambda_p+2\mu_p=0 \quad \Rightarrow \quad \nu_p=\lambda_p-2\mu_p.$$

In the following we reformulate the general set of equations
\eqref{Einstein-00pert}, \eqref{Einstein-11pert},
\eqref{lambda_pert}, \eqref{conserv_osc_pert_gen}, and \eqref{phi_pert_gen}
with this choice of gauge,
which allows us to eliminate the perturbation $\nu_p$ from the set of equations.
Employing again the dimensionless variables \eqref{dimless_xi_v},
Eq.~\eqref{phi_pert_gen} yields
\begin{equation}
\label{phi_pert_gen_lam0}
\phi_p^{\prime\prime}+
\frac{1}{2}\left(\nu_0^\prime+2\mu_0^\prime\right)\phi_p^\prime
+\omega^2 e^{-\nu_0}\phi_p=-4\Lambda^2 \tilde{V}(0)\left[\left(1-3\Lambda^2\phi_0^2\right)\phi_p+
\left(1-\Lambda^2\phi_0^2\right)\phi_0\lambda_p\right],
\end{equation}
where we have introduced a new dimensionless frequency $\bar{\omega}=\omega L$,
and subsequently again dropped the bar for notational simplicity.

The equations \eqref{Einstein-00pert} and \eqref{Einstein-22pert} give, respectively,
\begin{eqnarray}
\label{Einstein-00pert_lam0}
&&\mu_p^{\prime\prime}+\frac{1}{2}\mu_0^\prime\left(3\mu_p^\prime-\lambda_p^\prime\right)
-\left(\mu_0^{\prime\prime}+\frac{3}{4}\mu_0^{\prime 2}\right)\lambda_p
+e^{-\mu_0}\mu_p
\nonumber \\
&&=
-\left\{
n B \left[\frac{1}{\theta_0}+\sigma(n+1)\right]\theta_0^n\theta_p-\phi_0^\prime\left(\phi_p^\prime-\frac{1}{2}\phi_0^\prime \lambda_p\right)+
4\Lambda^2 \tilde{V}(0)\left(1-\Lambda^2\phi_0^2\right)\phi_0\phi_p
\right\},
\\
\label{Einstein-22pert_lam0}
&&\lambda_p^{\prime\prime}-\mu_p^{\prime\prime}
+\frac{1}{2}\nu_0^\prime\left(\lambda_p^\prime-3\mu_p^\prime\right)
-\lambda_p\left[
\mu_0^{\prime\prime}+\nu_0^{\prime\prime}+\frac{1}{2}\left(\mu_0^{\prime 2}+\nu_0^{\prime 2}+\mu_0^\prime \nu_0^\prime\right)
\right]
+\omega^2e^{-\nu_0}(\mu_p+\lambda_p) \nonumber
\\
&&=
-2\left[-B\sigma(n+1)\theta_0^n \theta_p-\phi_0^\prime\left(\phi_p^\prime-\frac{1}{2}\phi_0^\prime \lambda_p\right)+
4\Lambda^2 \tilde{V}(0)\left(1-\Lambda^2\phi_0^2\right)\phi_0\phi_p\right],
\end{eqnarray}
 while Eq.~\eqref{conserv_osc_pert_gen} leads to
\begin{align}
\label{conserv_osc_pert_mex}
\begin{split}
&\omega^2 e^{-\nu_0}\Big\{B\theta_0^n\left[1+\sigma(n+1)\theta_0\right] \psi+\phi_0^\prime\phi_p\Big\}-
B\sigma(n+1)\frac{d}{d\xi}\Big(\theta_0^n \theta_p\Big)
+\frac{d \tilde{V}_p}{d \xi}
 \\
&+\phi_0^{\prime\prime}\left(\phi_p^\prime-\frac{1}{2}\phi_0^\prime \lambda_p\right)
+\phi_0^\prime\left[\phi_p^{\prime\prime}-\frac{1}{2}\left(\phi_0^{\prime\prime}\lambda_p+
\phi_0^\prime \lambda_p^\prime\right)\right]
-\frac{1}{2}B \theta_0^n\left[\frac{n}{\theta_0}+\sigma(n+1)^2\right]\theta_p\nu_0^\prime
\\
&-\frac{1}{2}B \theta_0^n\left[1+\sigma(n+1)\theta_0\right](\lambda_p^\prime-2\mu_p^\prime)
+\frac{1}{2}\phi_0^{\prime 2}(\lambda_p^\prime-2\mu_p^\prime)
\\
&+\nu_0^\prime\phi_0^\prime\left(\phi_p^\prime-\frac{1}{2}\phi_0^\prime \lambda_p\right)+
\phi_0^\prime\left[\phi_0^\prime\mu_p^\prime+2\mu_0^\prime\left(\phi_p^\prime-\frac{1}{2}\phi_0^\prime \lambda_p\right)\right]=0,
\end{split}
\end{align}
where
$$
\frac{d \tilde{V}_p}{d \xi}=4\Lambda^2 \tilde{V}(0)
\left[\left(1-3\Lambda^2\phi_0^2\right)\phi_0^\prime\phi_p+\left(1-\Lambda^2\phi_0^2\right)\phi_0\phi_p^\prime
\right].
$$
Note, that we can eliminate $\psi$ in Eq.~\eqref{conserv_osc_pert_mex}
by replacing the curly brackets by the corresponding expression from
the constraint equation,
\begin{equation}
\label{lambda_pert_lam0}
2\mu_p^\prime +\left(\mu_0^\prime-\nu_0^\prime\right)\mu_p-\mu_0^\prime\lambda_p
-2\left\{B\theta_0^n\left[1+\sigma(n+1)\theta_0\right] \psi+\phi_0^\prime\phi_p\right\}=0,
\end{equation}
which results from Eq.~\eqref{lambda_pert}.
Here $\psi=\zeta/L$ is the dimensionless Lagrangian displacement.
The second constraint equation follows from Eq.~\eqref{Einstein-11pert} and reads
\begin{eqnarray}
\label{Einstein-11pert_lam0}
&&\mu_0^\prime(\lambda_p^\prime -\mu_p^\prime)+\nu_0^\prime \mu_p^\prime
-\lambda_p\left(\frac{1}{2}\mu_0^{\prime 2}+\mu_0^\prime \nu_0^\prime\right)
+2\left(\omega^2 e^{-\nu_0}+e^{-\mu_0}\right)\mu_p
\nonumber \\
&&=
-2\left[
- B \sigma(n+1)\theta_0^n\theta_p+\phi_0^\prime\left(\phi_p^\prime-\frac{1}{2}\phi_0^\prime \lambda_p\right)+
4\Lambda^2 \tilde{V}(0)\left(1-\Lambda^2\phi_0^2\right)\phi_0\phi_p
\right].
\end{eqnarray}

Thus, for the four functions $\phi_p, \lambda_p, \mu_p, \theta_p$,
we have the set of four equations
\eqref{phi_pert_gen_lam0}-\eqref{conserv_osc_pert_mex},
to investigate the stability of the configurations.
For this set of equations, we choose the following boundary conditions
at $\xi=0$:
\begin{equation}
\label{bound_cond_pert_mex}
\lambda_p(0)=\lambda_{p 0}, \quad
\mu_p(0)=\mu_{p 0},\quad
\theta_p(0)=\theta_{p 0}, \quad
\phi_p(0)=0,\quad \phi_p^\prime(0)=\phi_{p 1},
\end{equation}
where $\lambda_p, \mu_p, \theta_p$ are even functions,
while $\phi_p$ is an odd function. The value of $\phi_{p 1}$
is obtained from Eq.~\eqref{Einstein-11pert_lam0},
$$
\phi_{p 1}=B\sigma (n+1)\theta_{p 0}+\frac{1}{2}\lambda_{p 0}-\left(\omega^2 e^{-\nu_c}+\frac{1}{\Sigma_c^2}\right)\mu_{p 0}.
$$

Thus the system contains three parameters:
$\lambda_{p 0}, \mu_{p 0}$, and $\theta_{p 0}$.
Their values are chosen such that the following conditions are satisfied.
(i) At the boundary of the fluid, $\xi=\xi_b$,
the value of $\theta_p$ should remain finite to ensure
that $p_p$ [cf. Eq.~\eqref{pres_energ_pert}] meets the condition $p_p=0$
at the boundary \{see, e.g., Eq.~(60) in Ref.~\cite{Chandrasekhar:1964zz}\}.
(ii) Asymptotically,
as $\xi\to \pm \infty$, the perturbations $\lambda_p, \mu_p, \phi_p$ should tend to zero.
In this connection
it is useful to determine the asymptotic behavior of the solutions.
This can be given in analytic form.

\newpage
\medskip
\noindent (A): {\it Static solutions}:
\medskip

 $\left\{  \begin{tabular}{l}
$\phi_0\to 1/\Lambda-C_1 \exp{\left[-\sqrt{8\Lambda^2 \tilde{V}(0)}\,\xi\right]}\Big/\xi$;\\[\medskipamount]
$\Sigma_0 \to \xi$, \quad $\Sigma_0^\prime \to 1-C_2/\xi$;\\[\medskipamount]
$e^{\nu_0}\to 1-2\, C_2/\xi$.\\[\medskipamount]
\end{tabular}  \right.  $

\medskip
\noindent (B): {\it Perturbations}:
\medskip

$\left\{ \begin{tabular}{l}
$\phi_p\to C_3 \exp{\left(-\sqrt{-\beta^2}\xi\right)}\Big/\xi$;\\[\medskipamount]
$\mu_p \to C_4  \exp{\left(-\sqrt{-\omega^2}\xi\right)}$;\\[\medskipamount]
$\lambda_p \to -C_4  \sqrt{-\omega^2}\xi \exp{\left(-\sqrt{-\omega^2}\xi\right)}$.\\[\medskipamount]
\end{tabular}\right.  $

\noindent Here  $\beta^2=\omega^2-8\Lambda^2 \tilde{V}(0)$,
and the $C_i$ are integration constants. Hence,
to obtain decaying solutions for the perturbation $\phi_p$ of the scalar field,
$\beta^2$ should be negative.
Taking into account that in our case $\tilde{V}(0)$ is always positive,
this means that the condition $\omega^2<8\Lambda^2 \tilde{V}(0)$
should be satisfied.

\begin{figure}[t]
\centering
  \includegraphics[height=10.cm]{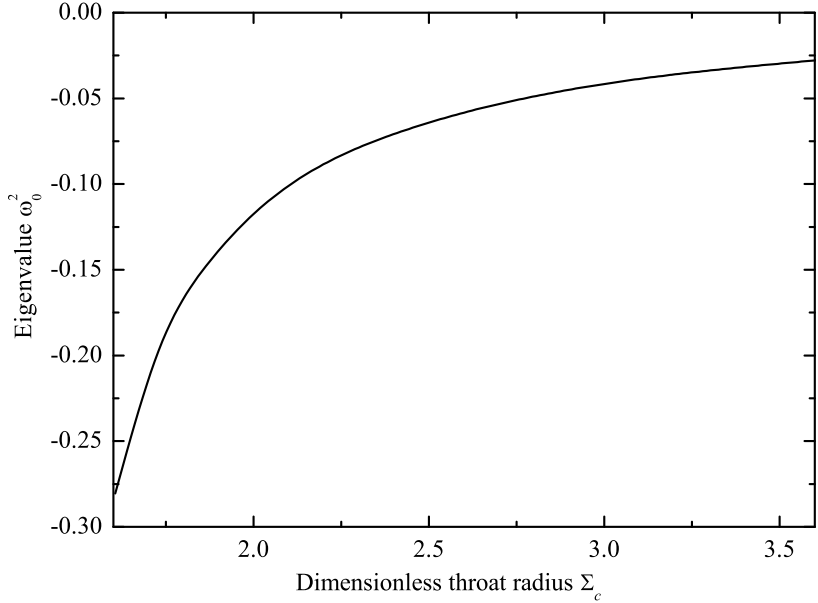}
\vspace{-1.cm}
\caption{The case without the fluid:
The  lowest eigenvalue $\omega_0^2$ is shown as a function of
the throat radius $\Sigma_c$.
As $\Sigma_c$ grows, $\omega^2 \to -0$.
}
\label{omega_Sigma_c}
\end{figure}

\begin{figure}[t]
\centering
  \includegraphics[height=10.cm]{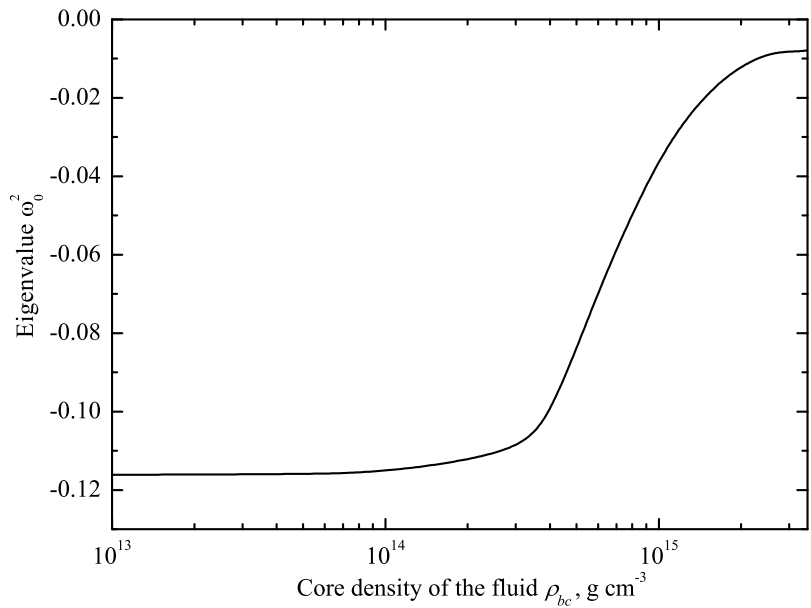}
\vspace{-1.cm}
\caption{The case with the fluid:
The  lowest eigenvalue $\omega_0^2$ is shown as a function of
the core density $\rho_{b c}$
(for $L=10\, \text{km}$ and $\Lambda=1$).
When $\rho_{b c}=B=0$ (i.e., without the fluid)
$\omega_0^2 \approx -0.11610638$.
}
\label{omega_rho_c}
\end{figure}

Let us now move to the results of the numerical calculations.
The set of equations \eqref{phi_pert_gen_lam0}-\eqref{conserv_osc_pert_mex}
together with the boundary conditions \eqref{bound_cond_pert_mex}
defines an eigenvalue problem for $\omega^2$.
The question of stability is thus reduced to a study of the possible
values of $\omega^2$.
If any of the values of $\omega^2$ are found to be negative,
then the perturbations will grow and the
configurations in question will be unstable against radial oscillations.

The results of the calculation of the eigenvalue
$\omega_0^2$ are shown in Figs.~\ref{omega_Sigma_c} and  \ref{omega_rho_c}. Figure~\ref{omega_Sigma_c}  shows the dependence of $\omega_0^2$
on the dimensionless throat radius $\Sigma_c$ of the wormhole solutions
without ordinary matter (i.e., with parameter $B=0$).
This case corresponds to the problem considered in Ref.~\cite{Kodama:1979}.
We thus find, that the eigenvalue $\omega_0^2$ is negative
for any size of the throat.
All these wormhole solutions are thus linearly unstable.
This result is in contrast to the restricted previous analysis,
which did not allow for perturbations of the throat radius.

Figure~\ref{omega_rho_c} shows our results for the case when the system contains the fluid,
i.e., for the star-plus-wormhole systems.
Here the  eigenvalue $\omega_0^2$ is shown
as a function of the core density of the fluid $\rho_{b c}$.
As the background solutions we employ the static solutions obtained in Sec.~\ref{static_solutions_mex}.
The initial value  $\mu_{p}(0)$ in Eq.~\eqref{bound_cond_pert_mex}
is chosen to be $\mu_{p 0}=1$,
and the values $\lambda_{p 0}$ and $\theta_{p 0}$ are chosen in such a way
that the solutions exhibit the asymptotic behavior shown in (B).
It is seen from Fig.~\ref{omega_rho_c} that
the square of the eigenfrequency remains always negative, independent of $\rho_{b c}$.
Thus, unfortunately, the star-plus-wormhole systems
obtained with the ghost scalar field with a quartic potential
are always unstable against linear perturbations.

\section{Conclusion}
\label{conclus}

Here we continued our study of star-plus-wormhole systems, begun in Refs.~\cite{Dzhunushaliev:2011xx,Dzhunushaliev:2012ke}.
The star-plus-wormhole systems considered there
were supported by a massless scalar field.
In the present paper we considered the case when the nontrivial topology
is provided by a scalar field with a quartic potential.
This choice of potential was based on the hope
that it would allow for stable solutions.

Choosing a neutron fluid with a polytropic equation of state \eqref{eqs_NS_WH},
that is filling the wormhole,
we showed that regular solutions for such mixed systems exist.
The resulting neutron stars have a finite size,
as in the case of the mixed configurations with a massless scalar field \cite{Dzhunushaliev:2011xx,Dzhunushaliev:2012ke}.

For our convenience in performing the stability analysis of such star-plus-wormhole systems,
we chose the values of the parameters in such a way
that the main mass of the configurations (more than 99\%)
is concentrated within the radius corresponding to the edge of the fluid.
In this case we showed the following.

\vspace{-0.2cm}
\begin{enumerate}
\itemsep=-0.2pt
\item[(1)] 
There exist regular static solutions found numerically by solving the coupled Einstein-matter
equations subject to a set of appropriate boundary conditions.
\item[(2)] 
Analyzing the dependence of the total mass on the core density of the fluid
(see Fig.~\ref{mass_rho_c}) and the mass-radius relation (see Fig.~\ref{M_R_relation}),
we found a range of core densities where stable configurations seemed to be possible.
\item[(3)] 
The linear stability analysis performed in Sec.~\ref{stab_anal_mex}, however,
indicated that the square of the lowest eigenfrequency of the perturbations
is always negative.
\end{enumerate}
This means that all star-plus-wormhole systems considered here
are unstable against linear perturbations.
Moreover, the wormholes themselves (i.e., the solutions without ordinary matter)
are unstable as well.
We conclude, that the star-plus-wormhole systems inherit their instability
from the wormholes.

In order to find stable star-plus-wormhole systems one should thus start
from stable wormholes. A possibility here would be to go beyond Einstein gravity
and include higher-curvature corrections
\cite{Kanti:2011jz,Kanti:2011yv}.
Such solutions would not need any exotic matter for their existence.

In the static solutions considered here the ghost scalar field tends
asymptotically in each of the universes to a different vacuum value,
given by the two degenerate minima of the potential.
The scalar field has therefore the shape of a kink \cite{rajaraman},
when considered as a function of the radial coordinate.
A similar behavior is found in the case of the
star-plus-wormhole configurations
considered in Refs.~\cite{Dzhunushaliev:2011xx,Dzhunushaliev:2012ke},
where asymptotically the massless ghost scalar field assumes values
that are equal in magnitude but have opposite signs.

Another interesting possibility consists in considering wormhole solutions,
which are supported by two interacting ghost scalar fields.
An example of such solutions was given in Ref.~\cite{Dzhunushaliev:2007cs}.
Here the scalar fields assume the same value asymptotically in both universes.
In our future work, we plan to perform a  stability analysis of those solutions.

\section*{Acknowledgement}

We gratefully acknowledge support by the Volkswagen Foundation and
by the German Research Foundation
within the framework of the DFG Research Training Group 1620 {\it Models of gravity}.
This work is  supported by the grant No.~514 in fundamental research in natural sciences
by the Ministry of Education and Science of Kazakhstan.


\begin{thebibliography}{999}

\bibitem{Wheeler:1955zz}
  J.~A.~Wheeler,
  Phys.\ Rev.\  {\bf 97}, 511 (1955).

\bibitem{Thorne:1988}
M.S. Morris and K.S. Thorne, Am. J. Phys. {\bf 56},  395 (1988);
M.S. Morris, K.S. Thorne, and U. Yurtsever, Phys. Rev. Lett. {\bf 61}, 1446 (1988).

\bibitem{Tonry:2003zg}
  J.~L.~Tonry {\it et al.}  [Supernova Search Team Collaboration],
  Astrophys.\ J.\  {\bf 594}, 1 (2003)
  [arXiv:astro-ph/0305008].

\bibitem{Alam:2003fg}
  U.~Alam, V.~Sahni, T.~D.~Saini, and A.~A.~Starobinsky,
  Mon.\ Not.\ R.\ Astron.\ Soc.\  {\bf 354}, 275 (2004)
  [arXiv:astro-ph/0311364].

\bibitem{Sullivan:2011kv}
  M.~Sullivan
  {\it et al.},
  Astrophys.\ J.\  {\bf 737}, 102 (2011)
  [arXiv:1104.1444 [astro-ph.CO]].


\bibitem{Mazur:2004ku}
  P.~O.~Mazur and E.~Mottola,
  ``Dark energy and condensate stars: Casimir energy in the large,''
  arXiv:gr-qc/0405111.
\bibitem{Dymnikova:2004qg}
  I.~Dymnikova and E.~Galaktionov,
  Classical Quantum Gravity  {\bf 22}, 2331 (2005)
  [arXiv:gr-qc/0409049].
\bibitem{Lobo:2005uf}
  F.~S.~N.~Lobo,
  Classical Quantum Gravity  {\bf 23}, 1525 (2006)
  [arXiv:gr-qc/0508115].
\bibitem{DeBenedictis:2005vp}
  A.~DeBenedictis, D.~Horvat, S.~Ilijic, S.~Kloster, and K.~S.~Viswanathan,
 Classical Quantum Gravity  {\bf 23}, 2303 (2006)
  [arXiv:gr-qc/0511097].
\bibitem{DeBenedictis:2008qm}
  A.~DeBenedictis, R.~Garattini, and F.~S.~N.~Lobo,
  Phys.\ Rev.\  D {\bf 78}, 104003 (2008)
  [arXiv:0808.0839 [gr-qc]].
\bibitem{Gorini:2008zj}
  V.~Gorini, U.~Moschella, A.~Y.~Kamenshchik, V.~Pasquier, and A.~A.~Starobinsky,
  Phys.\ Rev.\  D {\bf 78}, 064064 (2008)
  [arXiv:0807.2740 [astro-ph]].
\bibitem{Dzhunushaliev:2008bq}
  V.~Dzhunushaliev, V.~Folomeev, R.~Myrzakulov, and D.~Singleton,
  J. High Energy Phys. {\bf 07}  (2008) 094
  [arXiv:0805.3211 [gr-qc]].
\bibitem{Gorini:2009em}
  V.~Gorini, A.~Y.~Kamenshchik, U.~Moschella, O.~F.~Piattella, and A.~A.~Starobinsky,
  Phys.\ Rev.\  D {\bf 80}, 104038 (2009)
  [arXiv:0909.0866 [gr-qc]].
\bibitem{Dzhunushaliev:2011ma}
  V.~Dzhunushaliev, V.~Folomeev, and D.~Singleton,
  Phys. Rev. D {\bf 84}, 084025 (2011)
  [arXiv:1106.1267 [astro-ph.SR]].
\bibitem{Folomeev:2011aa}
 V.~Folomeev and D.~Singleton,
  Phys.\ Rev.\ D {\bf 85}, 064045 (2012)
  [arXiv:1112.1786 [astro-ph.SR]].

\bibitem{kuhfittig}
P.~K.~F.~Kuhfittig,
  Adv.\ Stud.\ Theor.\ Phys.\  {\bf 5}, 365 (2011)
  [arXiv:1001.0381 [gr-qc]].

\bibitem{lxli} L. X. Li, J. Geom. Phys. {\bf 40}, 154 (2001)

\bibitem{ArmendarizPicon:2002km}
  C.~Armendariz-Picon,
  Phys.\ Rev.\  D {\bf 65}, 104010 (2002)
  [arXiv:gr-qc/0201027].


\bibitem{Sushkov:2002ef}
  S.~V.~Sushkov and S.~W.~Kim,
 Classical Quantum Gravity  {\bf 19}, 4909 (2002)
  [arXiv:gr-qc/0208069].

\bibitem{Lemos:2003jb}
  J.~P.~S.~Lemos, F.~S.~N.~Lobo, and S.~Q. de Oliveira,
  Phys.\ Rev.\ D {\bf 68}, 064004 (2003)
  [gr-qc/0302049].

\bibitem{lobo}
F. S. N. Lobo, Phys. Rev. D {\bf 71}, 084011 (2005).

\bibitem{sushkov}
S. V. Sushkov, Phys. Rev. D {\bf 71}, 043520 (2005).




\bibitem{Kardashev:2006nj}
  N.~S.~Kardashev, I.~D.~Novikov, and A.~A.~Shatskiy,
  Int.\ J.\ Mod.\ Phys.\  D {\bf 16}, 909 (2007)
  [arXiv:astro-ph/0610441].
\bibitem{Carroll:2003st}
  S.~M.~Carroll, M.~Hoffman, and M.~Trodden,
  Phys.\ Rev.\ D {\bf 68}, 023509 (2003)
  [astro-ph/0301273].
\bibitem{Copeland:2006wr}
  E.~J.~Copeland, M.~Sami, and S.~Tsujikawa,
  Int.\ J.\ Mod.\ Phys.\  D {\bf 15}, 1753 (2006)
  [arXiv:hep-th/0603057].
\bibitem{Bronnikov:1973fh}
  K.~A.~Bronnikov,
  Acta Phys.\ Polon.\  B {\bf 4}, 251 (1973).

\bibitem{Ellis:1973yv}
  H.~G.~Ellis,
  J.\ Math.\ Phys.\  {\bf 14}, 104 (1973).


\bibitem{Gonzalez:2008wd}
  J.~A.~Gonzalez, F.~S.~Guzman, and O.~Sarbach,
 Classical Quantum Gravity  {\bf 26}, 015010 (2009)
  [arXiv:0806.0608 [gr-qc]].


\bibitem{Bronnikov:2011if}
  K.~A.~Bronnikov, J.~C.~Fabris, and A.~Zhidenko,
  Eur.\ Phys.\ J.\ C {\bf 71}, 1791 (2011).


\bibitem{Gonzalez:2008xk}
  J.~A.~Gonzalez, F.~S.~Guzman, and O.~Sarbach,
 Classical Quantum Gravity  {\bf 26}, 015011 (2009)
  [arXiv:0806.1370 [gr-qc]].



\bibitem{bronn3}
 K.A. Bronnikov  and S. Grinyok,
 Grav.  Cosmol. {\bf 7}, 297 (2001).

\bibitem{bronn4}
 K.A. Bronnikov and S.V. Grinyok,
 Grav. Cosmol. {\bf 10}, 237 (2004).

\bibitem{Bronnikov:2002qx}
  K.~A.~Bronnikov and S.~Grinyok,
  arXiv:gr-qc/0205131.

\bibitem{Bronnikov:2005an}
  K.~A.~Bronnikov and S.~V.~Grinyok,
  Grav.\ Cosmol.\  {\bf 11}, 75 (2005)
  [arXiv:gr-qc/0509062].




\bibitem{Kodama:1978dw}
T.~Kodama,Phys.\ Rev.\  D {\bf 18}, 3529 (1978).

\bibitem{Kodama:1979}
T. Kodama, L.C.S. de Oliveira, and F.C. Santos, Phys. Rev. D {\bf 19}, 3576 (1979).


\bibitem{Dzhunushaliev:2010bv}
  V.~Dzhunushaliev, V.~Folomeev, D.~Singleton, and R.~Myrzakulov,
  Phys.\ Rev.\  D {\bf 82}, 045032 (2010)
  [arXiv:1006.1527 [gr-qc]].



\bibitem{Dzhunushaliev:2011xx}
  V.~Dzhunushaliev, V.~Folomeev, B.~Kleihaus, and J.~Kunz,
J. Cosmol. Astropart. Phys. {\bf 04}  (2011) 031
  [arXiv:1102.4454 [astro-ph.GA]].

\bibitem{Dzhunushaliev:2012ke}
  V.~Dzhunushaliev, V.~Folomeev, B.~Kleihaus, and J.~Kunz,
  Phys.\ Rev.\ D {\bf 85}, 124028 (2012)
  [arXiv:1203.3615 [gr-qc]].

\bibitem{Bronnikov:2012ch}
  K.~A.~Bronnikov, R.~A.~Konoplya, and A.~Zhidenko,
  Phys.\ Rev.\ D {\bf 86}, 024028 (2012)
  [arXiv:1205.2224 [gr-qc]].


\bibitem{Stanuk1964}
K.P. Stanukovich, Soviet Physics Doklady {\bf 9}, 63 (1964).

\bibitem{Stanuk}
K.P. Stanukovich,
{\it Unsteady Flows of Continuous Medium}
(Nauka, Moscow, 1971).

\bibitem{Oppen1939}
J.R. Oppenheimer and G.M. Volkoff,  Phys. Rev.  {\bf 55}, 374  (1939).

\bibitem{Cameron1959}
A.G.W. Cameron, Astrophys.\ J.\ {\bf 130}, 884  (1959).

\bibitem{DAl1985}
J. Diaz-Alonso and J.M. Iba\~{n}ez-Cabanell, Astrophys.\ J.\ {\bf 291}, 308  (1985).
\bibitem{Haensel:2004nu}
  P.~Haensel and A.~Y.~Potekhin,
  Astron.\ Astrophys.\  {\bf 428}, 191 (2004)
  [astro-ph/0408324].

\bibitem{Salg1994}
 M. Salgado, S. Bonazzola, E. Gourgoulhon, and P. Haensel, Astron. Astrophys. {\bf 291}, 155 (1994).
\bibitem{Zeld1961}
Ya. B. Zel'dovich, J. Exp. Theoret. Phys. {\bf 41}, 1609 (1961)
[Sov. Phys. JETP {\bf 14}, 1143 (1962)].


\bibitem{Zeld}
Ya. B. Zel'dovich and  I. D. Novikov,
{\it Stars and relativity}
(Dover, New York, 1996).

\bibitem{Tooper2}
  R.~Tooper,
  Astrophys.\ J.\  {\bf 142}, 1541 (1965).



\bibitem{Colpi:1986ye}
  M.~Colpi, S.~L.~Shapiro, and I.~Wasserman,
  Phys.\ Rev.\ Lett.\  {\bf 57}, 2485 (1986).

\bibitem{Gleiser:1988rq}
  M.~Gleiser,
  Phys.\ Rev.\ D {\bf 38}, 2376 (1988);
  {\bf 39}, 1257(E) (1989).

\bibitem{Shinkai:2002gv}
  H.~-a.~Shinkai and S.~A.~Hayward,
  Phys.\ Rev.\ D {\bf 66}, 044005 (2002)
  [gr-qc/0205041].


\bibitem{Chandrasekhar:1964zz}
  S.~Chandrasekhar,
  Astrophys.\ J.\  {\bf 140}, 417 (1964).

\bibitem{Kanti:2011jz}
  P.~Kanti, B.~Kleihaus, and J.~Kunz,
  Phys.\ Rev.\ Lett.\  {\bf 107}, 271101 (2011)
  [arXiv:1108.3003 [gr-qc]].

\bibitem{Kanti:2011yv}
  P.~Kanti, B.~Kleihaus, and J.~Kunz,
  Phys.\ Rev.\ D {\bf 85}, 044007 (2012) 
%
  [arXiv:1111.4049 [hep-th]].

\bibitem{rajaraman}
R. Rajaraman,
{\it  An Introduction to Solitons and Instantons in Quantum Field Theory}
 (North-Holland Publishing Company, Amsterdam, New York, Oxford, 1982).
\bibitem{Dzhunushaliev:2007cs}
  V.~Dzhunushaliev and V.~Folomeev,
  Int.\ J.\ Mod.\ Phys.\ D {\bf 17}, 2125 (2008)
  [arXiv:0711.2840 [gr-qc]].
\end{thebibliography}
\end{document}